\documentclass[showpacs,twocolumn,aip,apl,reprint]{revtex4-1}
\usepackage{amsmath}
\usepackage{amsfonts}
\usepackage{amssymb}
\usepackage{amsthm}
\usepackage{graphicx}
\usepackage{CJK}

\begin{document}
\title{General planar transverse domain walls realized by optimized transverse magnetic field pulses in magnetic biaxial nanowires}
\author{Mei Li}
\affiliation{School of Physics and Technology, Center for Electron Microscopy and MOE Key Laboratory of Artificial Micro- and Nano-structures, Wuhan University, Wuhan 430072, China}
\author{Jianbo Wang}
\email{wang@whu.edu.cn}
\affiliation{School of Physics and Technology, Center for Electron Microscopy and MOE Key Laboratory of Artificial Micro- and Nano-structures, Wuhan University, Wuhan 430072, China}
\author{Jie Lu}
\email{jlu@mail.hebtu.edu.cn}
\affiliation{College of Physics and Information Engineering,
Hebei Advanced Thin Films Laboratory, Hebei Normal University, Shijiazhuang 050024, China}
\date{\today}

\begin{abstract}
We report the realization of a planar transverse domain wall (TDW) with arbitrary tilting angle
in a magnetic biaxial nanowire under a transverse magnetic field (TMF) pulse
with fixed strength and optimized orientation profile.
We smooth any twisting in azimuthal angle plane of a TDW
and thus completely decouple the polar and azimuthal degrees of freedom.
The analytical differential equation that describes the polar angle distribution is then derived
and the resulting solution is not a Walker-ansatz form.
With this optimized TMF pulse comoving, the field-driven dynamics of the planar TDW
is investigated. It turns out the comoving TMF pulse increases the wall velocity under the same axial driving field.
These results will help to design a series of modern logic and memory nanodevices based on general planar TDWs.
\end{abstract}



\maketitle


Magnetic nanowire (NW) devices, such as the domain-wall (DW) logics\cite{Science_309_1688_2005},
racetrack memories\cite{Science_320_190_2008},
and shift registers\cite{Science_320_209_2008}, etc., have developed rapidly in the past decades.
Advances in manufacturing thinner NWs greatly improve the integration level of these devices and make them
quasi one-dimensional (1D) systems, in which transverse DWs (TDWs) dominate\cite{IEEE_Trans_Magn_33_4167_1997,JMMM_290_750_2005}.
TDWs can be driven to propagate along wire axis by
magnetic fields\cite{Walker,Science_284_468_1999,nmat_2_85_2003,nmat_4_741_2005},
spin-polarized currents\cite{Berger_PRB_1996,Slonczewski_JMMM_1996,PRL_92_077205_2004,PRL_96_197207_2006}
or temperature gradient\cite{PRL_99_066603_2007,PRL_110_177202_2013}, etc.
Among them, the field-driven case is the most basic.
The Walker's analysis\cite{Walker} based on the Landau-Lifshitz-Gilbert (LLG) equation\cite{LLG_equation}
indicates the crucial role of the transverse magnetic anisotropy of a NW, which leads to
the Walker limit separating two distinct propagation modes: traveling-wave
and reciprocating rotation. However, the TDW tilting attitude in both modes cannot be arbitrarily controlled.

To manipulate the TDW tilting angle, using a uniform transverse magnetic field (TMF) is the easiest way and has been intensively
studied\cite{Sobolev2,jlu_TMF_JAP,AGoussev_PRB,AGoussev_Royal,jlu_prb_2016}.
However, a uniform TMF induces a twisting in TDW azimuthal plane\cite{jlu_prb_2016}.
In many circumstances, a complete planar TDW at any tilting angle is necessary for engineering applications.
In this work, we smooth the TDW twisting by changing the TMFs from uniform to space-dependent.
We focus on the case where the TMF strength is fixed and its orientation is allowed to change freely.
For statics, we will provide an optimized TMF profile that maintains a planar TDW with arbitrary tilting angle.
For dynamics, a TDW carrying this TMF profile along with it will acquire higher velocity than
that under pure axial driving field.

The system is sketched in Fig. 1.
A head-to-head (HH) TDW with width $\Delta$ is nucleated in a thin enough magnetic NW with thickness $t$ and width $w$.
The $z$ axis is along wire axis, the $x$ axis is in the thickness direction and $\hat{y}=\hat{z}\times\hat{x}$.
The magnetization $\vec{M}(\vec{r})$ with constant magnitude $M_s$ is fully described
by its polar angle $\theta(\vec{r})$ and azimuthal angle $\phi(\vec{r})$.
A TMF profile with constant strength $H_{\perp}$ and tunable orientation angle $\Phi_{\perp}(z)$,
\begin{equation}\label{TMF_vec}
    \vec{H}_{\mathrm{TMF}}(z)=H_{\perp}[\cos\Phi_{\perp}(z),\sin\Phi_{\perp}(z),0],
\end{equation}
is applied across the whole NW.

\begin{figure}[htbp]
\centering
\scalebox{0.38}[0.38]{\includegraphics[angle=0]{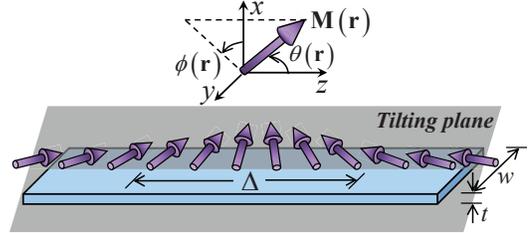}}
\caption{(Color online) A head-to-head TDW with width $\Delta$ in a nanowire with thickness $t$ and width $w$.
}\label{fig1}
\end{figure}

The time evolution of $\vec{M}(\vec{r})$ is described by the LLG equation,
\begin{equation}\label{LLG_vec}
\frac{\partial\vec{M}}{\partial t}=-\gamma\vec{M}\times\vec{H}_
{\mathrm{eff}}+\frac{\alpha}{M_s}\vec{M}\times\frac{\partial\vec{M}}{\partial t},
\end{equation}
where $\gamma$ is the gyromagnetic ratio, $\vec{H}_{\mathrm{eff}}$ is the effective field
$-(\partial E_{\mathrm{tot}}/\partial \vec{M})/\mu_{0}$.
For the NW under investigation, the total magnetic energy density is
\begin{eqnarray}\label{E_density}
E_{\mathrm{tot}}&=&\frac{J}{M_s^2}(\nabla\vec{M})^2-\frac{1}{2}k_1^0 \mu_0 M_z^2+\frac{1}{2}k_2^0 \mu_0 M_x^2 \nonumber \\
     & & +E_{\mathrm{m}}(\vec{M})-\mu_0\vec{M}\cdot(\vec{H}_{\parallel}+\vec{H}_{\mathrm{TMF}}),
\end{eqnarray}
where $J$ is the exchange coefficient, $k^0_{1(2)}$ is the crystalline anisotropy in the easy (hard) axis,
$\vec{H}_{\parallel}=H_1 \hat{e}_z$ is the axial driving field, and $E_{\mathrm{m}}$ is the magnetostatic energy density.

In thin enough NWs, by means of the ``nonlocal to local" simplification\cite{jlu_prb_2016},
most of $E_{\mathrm{m}}$ can be described by quadratic terms of $M_{x,y,z}$ in terms of three average demagnetization
factors $D_{x,y,z}$, thus $k_1^0\rightarrow k_1=k_1^0+(D_y-D_z)$ and $k_2^0\rightarrow k_2=k_2^0+(D_x-D_y)$.
For 1D systems, $\theta(\vec{r})\equiv\theta(z),\,\phi(\vec{r})\equiv\phi(z)$ hence
$(\nabla\vec{M})^2 = (M'_x)^2+(M'_y)^2+(M'_z)^2=M_s^2[(\theta')^2+\sin^2\theta(\phi')^2]$
where a prime means spatial derivative to $z$.
In the absence of any external fields, the total magnetic energy is
\begin{eqnarray}\label{Total_energy_wire}
  \mathcal{E} &=& wt\int_{V}\Bigg\{J[(\theta')^2+\sin^2\theta(\phi')^2] \nonumber \\
    & &    +\frac{1}{2}\mu_0 M_s^2\sin^2\theta(k_1+k_2\cos^2\phi)\Bigg\}dz,
\end{eqnarray}
in which we have redefined the energy origin by dropping $\mu_0 k_1 M_s^2 V/2$ with $V$ being the wire volume.

To obtain a stable TDW, we need to minimize $\mathcal{E}$. First one should let $\phi\equiv (n+1/2)\pi$ to
eliminate $(\phi')^2$ and $\cos^2\phi$ terms. This leads to a planar TDW lying in the easy plane.
Then the wire will have minimum energy when
$\Delta_0\theta'\equiv\pm\sin\theta$ where $\Delta_0=\sqrt{\frac{2J}{\mu_0 k_1 M_s^2}}$ and
$+(-)$ means HH (tail-to-tail, TT) TDW.
The resulting profile is the well-known Walker's solution,
\begin{equation}\label{Walker_ansatz_static}
  \theta(z)=2\tan^{-1}e^{\pm \frac{z-z_0}{\Delta_0}},
\end{equation}
with $z_0$ being the TDW center.
In brief, for a thin enough biaxial NW, the stable TDW is a planar wall which lies in the easy $yz-$plane.

In this work, we aim to realize a general planar TDW with arbitrary tilting attitude.
To achieve this, we need a TMF to pull the azimuthal angle plane out of the easy plane.
However, a uniform TMF generally induces twisting around the TDW center\cite{jlu_prb_2016}.
To erase the twisting, we fix the TMF strength and allow it rotate freely
to look for an optimal profile that results in a planar TDW.

Rewrite the vectorial LLG equation (\ref{LLG_vec}) to scalar form,
\begin{subequations}\label{LLG_scalar_exact}
\begin{align}
  \dot{\theta}(1+\alpha^2)/\gamma = A-\alpha B,   \\
  \dot{\phi}(1+\alpha^2)\sin\theta/\gamma =  B+\alpha A,
\end{align}
\end{subequations}
with
\begin{subequations}\label{AB_exact}
\begin{align}
  A &\equiv  -H_{\perp}\sin[\phi-\Phi_{\perp}(z)]+k_2 M_s \sin\theta\sin\phi\cos\phi \nonumber \\
    & \quad   +\frac{2J}{\mu_0 M_s\sin\theta}(\phi'\sin^2\theta)',   \\
  B &\equiv  H_1\sin\theta-H_{\perp}\cos\theta\cos[\phi-\Phi_{\perp}(z)] \nonumber \\
    & \quad   +M_s\sin\theta\cos\theta(k_1+k_2\cos^2\phi)  \nonumber \\
    &  \quad  -\frac{2J}{\mu_0 M_s}\left[ \theta''-(\phi')^2\sin\theta\cos\theta \right],
\end{align}
\end{subequations}
where a dot means time derivative.
To realize a static planar TDW, first we need the magnetization orientations in the two faraway domains.
In the left domain ($z\rightarrow-\infty$), the polar (azimuthal) angle of magnetization is denoted as $\theta_{\infty}$
($\phi_{\infty}$), while those in the right domain ($z\rightarrow+\infty$) are $\pi-\theta_{\infty}$ and $\phi_{\infty}$, respectively.
The static condition $ \dot{\theta}_{\infty}=0, \dot{\phi}_{\infty}=0$
and domain condition $\theta'_{\infty}=\theta''_{\infty}=0, \phi'_{\infty}=\phi''_{\infty}=0$
turn Eq. (\ref{LLG_scalar_exact}) to
\begin{subequations}\label{AB_static_in_domains}
\begin{align}
 H_{\perp}\sin[\phi_{\infty}-\Phi_{\perp}(z)] &= k_2 M_s \sin\theta_{\infty}\sin\phi_{\infty}\cos\phi_{\infty},  \\
H_{\perp}\cos[\phi_{\infty}-\Phi_{\perp}(z)] &= M_s\sin\theta_{\infty}(k_1+k_2\cos^2\phi_{\infty}).
\end{align}
\end{subequations}
The solution of Eq. (\ref{AB_static_in_domains}) is (we focus on HH TDWs)
\begin{subequations}\label{theta_phi_in_domains}
\begin{align}
    \theta_{\infty} & =\sin^{-1}\left(H_{\perp}/H_{\perp}^{\mathrm{max}}(z)\right),    \\
    \phi_{\infty}   & =\tan^{-1}\left[(1+k_2/k_1)\tan\Phi_{\perp}(z)\right],
\end{align}
\end{subequations}
with
\begin{equation}\label{H_perp_max}
  H_{\perp}^{\mathrm{max}}(z) = k_1 M_s\left[1+\frac{k_2(2k_1+k_2)}{k_1^2+(k_1+k_2)^2\tan^2\Phi_{\perp}(z)}\right]^{\frac{1}{2}}.
\end{equation}

From Eq. (\ref{theta_phi_in_domains}b), a necessary condition of the TDW being planar is $\Phi_{\perp}(z=\pm\infty)=\Phi_{\perp}^{\infty}$.
Without losing generality, suppose $0<\Phi_{\perp}^{\infty}<\pi/2$, we have $0<\Phi_{\perp}^{\infty}<\phi_{\infty}<\pi/2$.
In addition, the TDW existence condition $ \theta_{\infty}\neq \pi/2$ sets an upper limit of the TMF strength,
\begin{equation}\label{TMF_upper}
H_{\perp}<H_{\perp}^{\mathrm{max}}(\infty).
\end{equation}

Next we move to the TDW region. The static condition $A=B=0$ becomes
\begin{subequations}\label{AB_to_f1f2}
\begin{align}
  0 &= f_1(\theta,\phi),  \\
  \frac{2J}{\mu_0 M_s}\theta'' &= f_2(\theta,\phi)\cos\theta + M_s(k_1+k_2\cos^2\phi_{\infty})\cos\theta\cdot \nonumber \\
    & \quad \left\{\sin\theta-\frac{H_{\perp}\cos[\phi_{\infty}-\Phi_{\perp}(z)]}{M_s(k_1+k_2\cos^2\phi_{\infty})}\right\},
\end{align}
\end{subequations}
with
\begin{subequations}\label{f1f2_definition}
\begin{align}
  f_1(\theta,\phi) &\equiv -H_{\perp}\sin[\phi-\Phi_{\perp}(z)]+k_2M_s\sin\theta\sin\phi\cos\phi \nonumber \\
      & \quad  +\frac{2J}{\mu_0 M_s}(2\theta'\phi'\cos\theta+\phi''\sin\theta),  \\
 f_2(\theta,\phi) &\equiv \frac{2J}{\mu_0 M_s}(\phi')^2\sin\theta+k_2 M_s\sin\theta(\cos^2\phi-\cos^2\phi_{\infty})  \nonumber \\
     & \quad +H_{\perp}\left\{\cos[\phi_{\infty}-\Phi_{\perp}(z)]-\cos[\phi-\Phi_{\perp}(z)]\right\}.
\end{align}
\end{subequations}

Now we consider a planar TDW
\begin{equation}\label{TDW_planar_phi}
\phi(z)\equiv \phi_{\infty}.
\end{equation}
Obviously, this solution makes $f_2(\theta,\phi)=0$ and thus
\begin{eqnarray}\label{TDW_static_theta_1}
\frac{2J}{\mu_0 M_s}\theta'' &=& M_s(k_1+k_2\cos^2\phi_{\infty})\cos\theta\cdot \nonumber \\
      & &\left\{\sin\theta-\frac{\cos[\phi_{\infty}-\Phi_{\perp}(z)]}{\cos(\phi_{\infty}-\Phi_{\perp}^{\infty})}\sin\theta_{\infty}\right\}.
\end{eqnarray}
On the other hand, $f_1(\theta,\phi)=0$ is reduced to
\begin{equation}\label{TDW_static_f1_new}
 H_{\perp}\sin[\phi_{\infty}-\Phi_{\perp}(z)] = k_2 M_s \sin\theta\sin\phi_{\infty}\cos\phi_{\infty}.
\end{equation}
Compare Eq. (\ref{TDW_static_f1_new}) with Eq. (\ref{AB_static_in_domains}a), we obtain
the dependence of TMF orientation on TDW polar angle,
\begin{equation}\label{TDW_static_TMF_profile_2}
\Phi_{\perp}(z) = \phi_{\infty}- \sin^{-1}\left[\frac{\sin\theta(z)}{\sin\theta_{\infty}}
           \cdot \sin(\phi_{\infty}-\Phi_{\perp}^{\infty})\right].
\end{equation}
or vice versa. Eq. (\ref{TDW_static_TMF_profile_2}) shows that the TMF cannot be uniform.
It also requires $|\sin\theta(z)\sin(\phi_{\infty}-\Phi_{\perp}^{\infty})/\sin\theta_{\infty}|\leq 1$,
which sets a lower limit of the TMF strength,
\begin{equation}\label{TMF_lower}
H_{\perp}\geq H_{\perp}^{\mathrm{min}}=\frac{k_1}{\sqrt{k_1^2+(k_1+k_2)^2\tan^2\Phi_{\perp}^{\infty}}}\cdot H_{\perp}^{\mathrm{max}}(\infty).
\end{equation}

Put Eq. (\ref{TDW_static_TMF_profile_2}) back into Eq. (\ref{TDW_static_theta_1}), we then have
\begin{eqnarray}\label{TDW_static_theta_2}
\Delta^2(\phi_{\infty})\theta'' &=& \left[\sin\theta-\sqrt{(1+\beta)\sin^2\theta_{\infty}-\beta\sin^2\theta}\right]\cos\theta,  \nonumber  \\
\beta  &=&   \tan^2(\phi_{\infty}-\Phi_{\perp}^{\infty}), \nonumber  \\
\Delta(\phi_{\infty}) & =&  \Delta_0 \left[1+(k_2/k_1)\cos^2\phi_{\infty}\right]^{-1/2},
\end{eqnarray}
where $\theta_{\infty}$ and $\phi_{\infty}$ are given
by Eq. (\ref{theta_phi_in_domains}) with $\Phi_{\perp}(z)\equiv\Phi_{\perp}^{\infty}$.

When $\Phi_{\perp}^{\infty}=n\pi/2$, $\phi_{\infty}\equiv\Phi_{\perp}^{\infty}$ hence $\beta=0$.
Eq. (\ref{TDW_static_theta_2}) is reduced to a Walker-ansatz-like form,
\begin{equation}\label{TDW_static_theta_3}
\Delta^2(\phi_{\infty})\theta''=\left(\sin\theta-\sin\theta_{\infty}\right)\cos\theta.
\end{equation}
Its solution has been presented by Eq. (16) in Ref. \cite{jlu_prb_2016}.

For $0<\Phi_{\perp}^{\infty}<\pi/2$, $\beta>0$ and Eq. (\ref{TDW_static_theta_2}) is not a
typical Walker-ansatz form. After some algebra, we have
\begin{eqnarray}\label{TDW_static_theta_4}
  g(\theta) &\equiv& \Delta^2(\phi_{\infty})(\theta')^2=\sin^2\theta-\sqrt{\beta}\sin\theta\sqrt{\eta^2-\sin^2\theta}  \nonumber \\
              & &     -\sqrt{\beta}\eta^2\left[\arcsin\frac{\sin\theta}{\eta}-\arcsin\sqrt{\beta/(1+\beta)}\right], \nonumber \\
         \eta &=&\sqrt{1+\beta^{-1}}\sin\theta_{\infty}.
\end{eqnarray}
When $z\geq z_0$, $\theta\geq\pi/2$. From Eq. (\ref{TDW_static_theta_4}), in principle the following integral
\begin{equation}\label{TDW_static_theta_5}
  \frac{z-z_0}{\Delta(\phi_{\infty})}=\int_{\pi/2}^{\theta}d\chi [g(\chi)]^{-\frac{1}{2}}\equiv g_1(\theta),
\end{equation}
gives the right-half profile $\theta(z)=g_1^{-1}(z-z_0)$.
For $z<z_0$, $\theta(z-z_0)=\pi-\theta(z_0-z)$.
Put it back into Eq. (\ref{TDW_static_TMF_profile_2}), the corresponding TDW orientation profile is obtained.

Based on the above analytics, we propose the algorithm of realizing an arbitrary planar TDW:
(1) Given TDW tilting angle $\phi_{\infty}$, Eq. (\ref{theta_phi_in_domains}b) gives $\Phi_{\perp}^{\infty}$ and hence $\beta$.
(2) Eqs. (\ref{TMF_upper}) and (\ref{TMF_lower}) provide $H_{\perp}^{\mathrm{max}}$ and $H_{\perp}^{\mathrm{min}}$.
(3) For an allowed $H_{\perp}$, Eq. (\ref{theta_phi_in_domains}a) gives $\theta_{\infty}$.
(4) Eq. (\ref{TDW_static_theta_5}) gives the $\theta$ profile.
(5) TMF follows Eq. (\ref{TDW_static_TMF_profile_2}).

We illustrate our algorithm in a $5\,\mathrm{nm}\times 100\,\mathrm{nm}\times 10\,\mathrm{\mu m}$ NW.
The results are shown in Fig. 2.
For this wire geometry, $D_x=0.92793,\, D_y=0.07140,\, D_z=0.00067$.
The magnetic parameters are: $M_s=500\,\mathrm{kA/m}$, $J=40\times10^{-12}\,\mathrm{J/m}$,
$k_1^0=4/\pi$, $k_2^0=0.3 k_1^0$, and $\alpha=0.1$.
Then $k_1=k_1^0+(D_y-D_z)=1.34397$ and $k_2=k_2^0+(D_x-D_y)=1.23851$.
We want to realize a planar TDW with tilting angle $\phi_{\infty}=5\pi/12$. Thus we must have $\Phi_{\perp}^{\infty}=0.34865\pi$,
hence $H_{\perp}^{\mathrm{min}}\approx 2375\,\mathrm{Oe}$ and $H_{\perp}^{\mathrm{max}}\approx 9174\,\mathrm{Oe}$.
We take $H_{\perp}=3\,\mathrm{kOe}$ as an example.
The TMF orientation profile is shown by black solid curve
and the resulting $\theta(\phi)$ profile of the planar TDW is indicated by red dashed curve (blue dotted line).
Obviously to maintain the planar TDW, the TMF should be closer to the hard $xz-$plane around the TDW center
to resist its twisting trend.

\begin{figure}[htbp]
\centering
\scalebox{0.88}[0.88]{\includegraphics[angle=0]{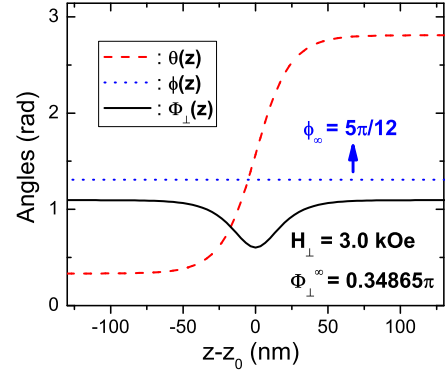}}
\caption{(Color online) Example of TMF orientation profile and its resulting
planar TDW in a $5\,\mathrm{nm}\times 100\,\mathrm{nm}\times 10\,\mathrm{\mu m}$ biaxial nanowire.
}\label{fig2}
\end{figure}

Next we turn to planar TDW dynamics.
From Eqs. (\ref{TMF_upper}) and (\ref{TMF_lower}), the TMF strength should be finite,
thus we rescale the axial driving field and the TDW propagation
velocity $V$ simultaneously\cite{AGoussev_PRB,AGoussev_Royal,jlu_prb_2016},
\begin{equation}\label{finiteTMF_scaling}
    H_1=\epsilon h_1,\quad V=\epsilon v,
\end{equation}
where $\epsilon$ is a dimensionless infinitesimal.
We focus on the traveling-wave mode and define the traveling coordinate
\begin{equation}\label{finiteTMF_xi}
    \xi\equiv z-V t=z-\epsilon v t.
\end{equation}
The TMF in dynamical case takes the same profile as that in static case, except for the
substitution $z\rightarrow \xi$ which means it moves along with the TDW center.
Then we expand $\theta(z,t)$, $\phi(z,t)$ as follows:
\begin{subequations}\label{finiteTMF_series_expansion}
\begin{align}
  \theta(z,t) &= \theta_0(\xi)+\epsilon\theta_1(\xi)+O(\epsilon^2), \\
  \phi(z,t) &= \phi_0(\xi)+\epsilon\phi_1(\xi)+O(\epsilon^2).
\end{align}
\end{subequations}
Put Eq. (\ref{finiteTMF_series_expansion}) into LLG equation (\ref{LLG_scalar_exact}),
to the zeroth order of $\epsilon$, we have
\begin{subequations}\label{finiteTMF_0order_equations_A0B0}
\begin{align}
  0 &=H_{\perp}\sin[\Phi_{\perp}(\xi)-\phi_0]+k_2 M_s \sin\theta_0\sin\phi_0\cos\phi_0 \nonumber \\
    & \quad  +\frac{2J}{\mu_0 M_s}(2\theta'_0\phi'_0\cos\theta_0+\phi''_0\sin\theta_0),  \\
  0 &=k_1 M_s\sin\theta_0\cos\theta_0\left[1+\frac{k_2}{k_1}\cos^2\phi_0+\frac{(\phi'_0)^2}{c^2}\right] \nonumber  \\
    & \quad  -H_{\perp}\cos\theta_0\cos[\Phi_{\perp}(\xi)-\phi_0]-\frac{2J}{\mu_0 M_s}\theta''_0,
\end{align}
\end{subequations}
where a prime means partial derivative with respect to $\xi$.
Under the comoving TMF profile (\ref{TDW_static_TMF_profile_2}) ($z\rightarrow \xi$),
the solution of Eq. (\ref{finiteTMF_0order_equations_A0B0}) is just Eqs. (\ref{TDW_planar_phi}) and (\ref{TDW_static_theta_5}).
To obtain the TDW velocity, we need to proceed to the next order.

At first order of $\epsilon$, we have
\begin{subequations}\label{finiteTMF_1order_equations_A1B1}
\begin{align}
  A_1 &= -v(\theta'_0+\alpha\sin\theta_0 \phi'_0)/\gamma,   \\
  B_1 &= -v(-\alpha\theta'_0+\sin\theta_0\phi'_0)/\gamma,
\end{align}
\end{subequations}
where
\begin{subequations}\label{finiteTMF_A1}
\begin{align}
  A_1 &= \mathbf{P}\theta_1+\mathbf{Q}\phi_1,     \\
  \mathbf{P} & \equiv \frac{2J}{\mu_0M_s}\left[2\phi'_0\left(\cos\theta_0\cdot\frac{\partial}{\partial\xi}-\theta'_0\sin\theta_0\right)
      +\phi''_0\cos\theta_0\right]  \nonumber \\
      & \quad +k_2 M_s\cos\theta_0\sin\phi_0\cos\phi_0, \\
  \mathbf{Q} &\equiv -H_{\perp}\cos[\Phi_{\perp}(\xi)-\phi_0]+k_2 M_s \sin\theta_0\cos 2\phi_0 \nonumber \\
      & \quad +\frac{2J}{\mu_0 M_s}\left(2\theta'_0\cos\theta_0\frac{\partial}{\partial\xi}+\sin\theta_0\frac{\partial^2}{\partial\xi^2}\right),
\end{align}
\end{subequations}
and
\begin{subequations}\label{finiteTMF_B1}
\begin{align}
 B_1 &= h_1\sin\theta_0+\mathbf{R}\theta_1+\mathbf{S}\phi_1,    \\
\mathbf{R} &\equiv k_1 M_s\cos 2\theta_0\left[1+\frac{k_2}{k_1}\cos^2\phi_0+\frac{(\phi'_0)^2}{c^2}\right] \nonumber \\
           & \quad +H_{\perp}\sin\theta_0\cos[\Phi_{\perp}(\xi)-\phi_0]-\frac{2J}{\mu_0 M_s}\frac{\partial^2}{\partial\xi^2},   \\
\mathbf{S} &\equiv k_1 M_s\sin 2\theta_0\left(\frac{\phi'_0}{c^2}\frac{\partial}{\partial\xi}-\frac{k_2}{k_1}\sin\phi_0\cos\phi_0\right)\nonumber \\
           & \quad  -H_{\perp}\cos\theta_0\sin[\Phi_{\perp}(\xi)-\phi_0].
\end{align}
\end{subequations}

We need to simplify $\mathbf{R}$ and $\mathbf{S}$ for $v(h_1)$ relationship.
It is clear that $\theta_0$ and $\phi_0$ have been fully decoupled. The partial derivative of ``$B_0=0$"
with respect to $\theta_0$ gives
\begin{eqnarray}\label{finiteTMF_B0eq0_theta0}
   \frac{2J}{\mu_0 M_s}\frac{\theta'''_0}{\theta'_0}
   &=& k_1 M_s\cos 2\theta_0\left[1+\frac{k_2}{k_1}\cos^2\phi_0+\frac{(\phi'_0)^2}{c^2}\right] \nonumber \\
   & & +H_{\perp}\sin\theta_0\cos[\Phi_{\perp}(z)-\phi_0],
\end{eqnarray}
hence simplifies $\mathbf{R}$ to
\begin{equation}\label{finiteTMF_R_simplify}
    \mathbf{R}=\frac{2J}{\mu_0 M_s}\left(-\frac{\partial^2}{\partial\xi^2}+\frac{\theta'''_0}{\theta'_0}\right),
\end{equation}
which is the same 1D self-adjoint Schr\"{o}dinger operator $\mathbf{L}$ as in Refs. \cite{AGoussev_PRB,AGoussev_Royal,jlu_prb_2016}.
Meantime, by partially differentiating $B_0=0$ with respect to $\phi_0$, we have
\begin{eqnarray}\label{finiteTMF_B0eq0_phi0}
0 &=& k_1 M_s\sin 2\theta_0\left(\frac{\phi''_0}{c^2}-\frac{k_2}{k_1}\sin\phi_0\cos\phi_0\right) \nonumber \\
  & & -H_{\perp}\cos\theta_0\sin[\Phi_{\perp}(\xi)-\phi_0],
\end{eqnarray}
which simplifies $\mathbf{S}$ to
\begin{equation}\label{finiteTMF_S_simplify}
\mathbf{S}=\frac{k_1 M_s \sin 2\theta_0}{c^2}\left(\phi'_0\frac{\partial}{\partial\xi}-\phi''_0\right)\equiv 0,
\end{equation}
since $\phi_0=const$. As a result, Eq. (\ref{finiteTMF_1order_equations_A1B1}b) becomes
\begin{equation}\label{finiteTMF_Ltheta1}
    \mathbf{L}\theta_1=-h_1\sin\theta_0+(-v)(-\alpha\theta'_0)/\gamma.
\end{equation}
The ``Fredholm alternative" requests the right hand side of Eq. (\ref{finiteTMF_Ltheta1})
to be orthogonal to the kernel of $\mathbf{L}$  (i.e., $\theta'_0$) for a solution $\theta_1$ to exist.
Noting that from Eq. (\ref{TDW_static_theta_4}),
\begin{equation}\label{theta_prime_square_inner_product}
\langle\theta'_0,\theta'_0\rangle<\Delta^{-1}(\phi_{\infty})\langle\theta'_0,\sin\theta_0\rangle,
\end{equation}
thus the planar TDW acquires a higher velocity than the Walker result,
\begin{equation}\label{finiteTMF_velocity}
    V_{\mathrm{Planar}}=\frac{\langle\theta'_0,\sin\theta_0\rangle}{\langle\theta'_0,\theta'_0\rangle}\frac{\gamma}{\alpha}H_1
    > \frac{\gamma\Delta(\phi_{\infty})}{\alpha}H_1.
\end{equation}

Finally, we would like to clarify that our strategy differs from that in Ref. \cite{PRL_104_037206_2010}, in which
they maximized the wall velocity by optimizing field pulses with fixed strength and totally free orientation.
In our work, we realize a planar TDW at any tilting angle by optimizing TMF pulses with fixed strength and tunable orientation.
The total external field also has fixed strength, but cannot freely orientate since it has a specified axial component.
In brief, our strategy is not optimal for the purpose of maximizing wall velocity.
However, it manipulates general planar TDWs which
should have widespread applications in modern nanodevice engineering.

This work is supported by the National Natural Science Foundation of China (Grants No. 11374088 and No. 51271134).

\end{document}